\documentstyle[aps,prl,twocolumn,epsf]{revtex}
\tighten
\newcommand{\etal}{\it et al. \rm}
\newcommand{\rt}{\rightarrow}
\newcommand{\pipi}{\pi^+ \pi^-}
\newcommand{\ppbar}{p \overline{p}}
\newcommand{\pbar}{\overline{p}}

\newcommand{\jpsi}{J/\psi}
\newcommand{\ee}{e^+ e^-}
\title{Observation of a near-threshold enhancement 
in the {\boldmath $\ppbar $} mass spectrum from 
radiative {\boldmath $\jpsi\rt\gamma\ppbar$} decays}

\author{(BES Collaboration)\\
J.~Z.~Bai$^1$,        Y.~Ban$^{9}$,          J.~G.~Bian$^1$,
X.~Cai$^{1}$,          J.~F.~Chang$^1$,
H.~F.~Chen$^{16}$,    H.~S.~Chen$^1$,
J.~Chen$^{3}$,        Jie~Chen$^{8}$,        J.~C.~Chen$^1$
% ...
}
\address{
$^1$ Institute of High Energy Physics, Beijing 100039, People's Republic of
     China\\
% ...
$^3$ Colorado State University, Fort Collins, Colorado 80523\\
% ...
$^8$ Nankai University, Tianjin 300071, People's Republic of China\\
$^{9}$ Peking University, Beijing 100871, People's Republic of China\\
% ...
$^{16}$ University of Science and Technology of China, Hefei 230026,
        People's Republic of China\\
% ...
}

\date{\today}
\begin{document}
\maketitle

\begin{abstract}
We observe a narrow enhancement near $2m_p$ in the invariant 
mass spectrum of $\ppbar$ pairs from radiative $\jpsi\rt\gamma\ppbar$
decays.   No similar structure is seen in $\jpsi\rt \pi^0\ppbar$ decays.
The results are based on an analysis of a 58 million event 
sample of $\jpsi$ decays accumulated with the BESII detector at the
Beijing electron-positron collider.  The enhancement can be fit with
either an $S$- or $P$-wave Breit Wigner resonance function.
In the case of the $S$-wave fit, the peak mass is below $2m_p$ at
$M = 1859 ^{~+3}_{-10}~{\rm (stat)} ^{~+5}_{-25}~{\rm (sys)}~{\rm 
MeV}/c^2$ and the total width is $\Gamma < 30 $~MeV/$c^2$ at the 90 
percent confidence level. These mass and width values
are not consistent with the properties of any known particle.
\end{abstract}
% insert suggested PACS numbers in braces on next line

%\vspace*{0.5cm}
%
% Begin main paper here
%%%%%%%%%%%%%%%%%%%%%%%%%%%%%%%%%%%%%%%%%%%%%%%%%%%%%%%%%%%%%%%%%%%%%%%%

There is an accumulation of evidence for anomalous behavior in the
proton-antiproton ($\ppbar$) system very near the $M_{\ppbar}=2m_p$
mass threshold.
The observed cross section for $\ee\rt hadrons$
has a narrow dip-like structure
at a center of mass energy of 
$\sqrt{s} \simeq 2m_p c^2$~\cite{FENICE}.
The proton's time-like magnetic
form-factor, determined from high statistics measurements of the
$\ppbar\rt\ee$ annihilation process,  exhibits a very steep fall-off
just above the $\ppbar$ mass threshold~\cite{LEAR}.
These data are suggestive of a narrow, $S$-wave triplet 
$\ppbar$ resonance with $J^{PC} = 1^{--}$ and mass near
$2m_p$.  In studies of $\pbar$ annihilations
at rest in deuterium, anomalies in the charged pion momentum 
spectrum from $\pbar d\rt \pi^-\pi^0 p$ and $\pi^+\pi^- n$ 
reactions~\cite{bridges} and the proton spectrum from 
$\pbar d\rt p 2\pi^+ 3\pi^-$~\cite{dalkarov} have
been interpreted as effects of narrow, below-threshold resonances.
There are no well established mesons that could
be associated with such states.  The proximity in mass
to $2m_p$ is suggestive of nucleon-antinucleon ($N\overline{N}$) 
bound states,  an idea that has a long history.
In 1949, Fermi and Yang~\cite{Fermi} proposed
that the pion was a tightly bound $N\overline{N}$ state.  
Nambu and Jona-Lasinio~\cite{Nambu} expanded on
this in 1961 with a model based on chiral 
symmetry that has, in addition to
a low-mass pion, a scalar $\ppbar$ composite state with mass equal
to $2m_p$.  Although these ideas have been
superseded by the quark model~\cite{kunihiro}, the possibility of
bound $N\overline{N}$ states with mass near $2m_p$,
generally referred to as {\it baryonium}, continues 
to be considered~\cite{richard}.
Recently
Belle has reported observations of the decays
$B^+\rt K^+\ppbar$~\cite{Belle1} and
$\overline{B}{}^0\rt D^0\ppbar$~\cite{Belle2}. In 
both processes there are enhancements
in the ${\ppbar}$ invariant mass distributions
near $M_{\ppbar}\simeq 2m_p$.
An investigation of low mass $\ppbar$
systems with different quantum numbers 
may help clarify the situation.

In this
letter we report a study of the low mass $\ppbar$ pairs
produced via radiative decays in a sample of 58 million $\jpsi$ 
events accumulated in the upgraded
Beijing Spectrometer (BESII) 
located at the  Beijing Electron-Positron Collider
(BEPC) at the Beijing Institute of
High Energy Physics.  For this reaction,
charge-parity conservation insures that 
the $\ppbar$ system has $C=+1$.

BESII is a large solid-angle magnetic spectrometer 
that is described in detail in ref.~\cite{BES}.  Charged
particle momenta are determined with a resolution of
$\sigma_p/p = 1.7\%\sqrt{1+p^2({\rm GeV}^2)}$ in a 40-layer cylindrical 
drift chamber.  Particle identification is accomplished
by specific ionization ($dE/dx$) measurements in the
drift chamber and time-of-flight (TOF) measurements in a 
barrel-like array of 48 scintillation counters.
The $dE/dx$ resolution is $\sigma_{dE/dx} = 8.4\%$;
the TOF resolution is $\sigma_{TOF}=180$~ps; 
both systems independently
provide more than $3\sigma$ separation of protons from any
other charged particle species
for the entire momentum range considered in this experiment. 
Radially outside of the time-of-flight counters is a 
12-radiation-length barrel shower counter (BSC) comprised of
gas proportional tubes interleaved with lead sheets.  The BSC
measures the energies and directions of photons with resolutions
of $\sigma_E/E\simeq 22\%/\sqrt{E({\rm GeV})}$, $\sigma_\phi=4.5$~mrad,
and $\sigma_\theta = 7.9$~mrad. The iron flux return of the
magnet is instrumented with three double layers of counters
that are used to identify muons.    

For this analysis we  use events with a high energy gamma ray
and two oppositely charged tracks each of which is well fitted
to a helix originating near the interaction point.
Candidate $\gamma$'s are
associated with energy clusters in the BSC that
have less than 80\% of their total energy in any one
readout layer and do not match the extrapolated
position of any charged track.   Since antiprotons that stop
in the material of the TOF or BSC can produce
annihilation products that are reconstructed
elsewhere in the detector as $\gamma$ rays,
no restrictions are placed on the total number
of neutral clusters in the event.
We use charged tracks and $\gamma$'s that are within the polar angle
region $|\cos\theta|<0.8$ and reject events where
both tracks are identified
as muons, or produce high energy showers in the BSC that are
characteristic of electrons.  
The $dE/dx$ information is used 
to form particle identification confidence levels
${\mathcal P}^i_{pid}$, where  $i$ denotes $\pi$, $K$ and $p$. 
We require that both charged tracks have 
${\mathcal P}^{p}_{pid} > {\mathcal P}^{K}_{pid} $
and  ${\mathcal P}^{p}_{pid} > {\mathcal P}^{\pi}_{pid} $.
A study based on a kinematically selected
sample of $\jpsi\rt K^{*\pm}K^{\mp}\rt K^+K^-\pi^0$ events indicate 
that the probability for a charged kaon to satisfy this requirement
is less than 1\% per track.

We subject the surviving events to four-constraint kinematic fits to the
hypotheses $\jpsi\rt\gamma\ppbar$ and $\jpsi\rt\gamma K^+K^-$.
For events with more that one $\gamma$, we 
select the $\gamma$ that has the highest fit confidence level.
We select events that have fit confidence level 
$CL_{\gamma\ppbar}>0.05$ and reject events that have
$CL_{\gamma K^+K^-} > CL_{\gamma\ppbar}$.  

\begin{figure}[htb]
\centerline{\epsfysize 1.8 truein
%\centerline{\epsfysize 1.6 truein
\epsfbox{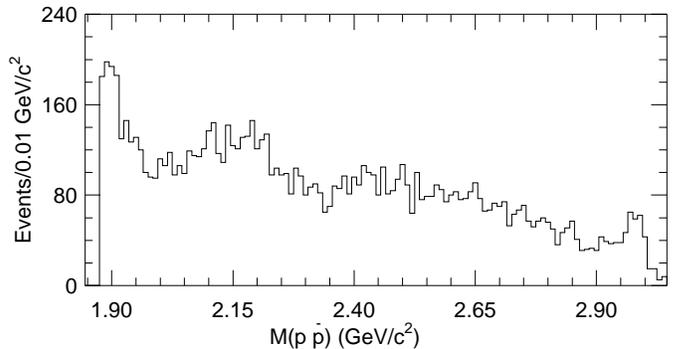}}
\caption{\label{fig:2pg_data_ihep}
The $\ppbar$ invariant mass distribution for the
$\jpsi\rt\gamma \ppbar$-enriched event sample.}
\end{figure}

Figure~\ref{fig:2pg_data_ihep} shows the $\ppbar$ invariant mass
distribution for surviving events.  The distribution has 
a peak near $M_{\ppbar}=2.98$~GeV/$c^2$ that is consistent
in mass, width, and yield with expectations for $\jpsi\rt\gamma\eta_c$,
$\eta_c\rt\ppbar$~\cite{dong},  a broad enhancement around 
$M_{\ppbar}\sim 2.2$~GeV/$c^2$,
and a narrow, low-mass peak at the $\ppbar$  mass
threshold that is the subject of this Letter.

Backgrounds from processes involving charged particles that are
not protons and antiprotons are negligibly small.
In addition to being well separated from other charged particles
by the $dE/dx$ measurements and the kinematic fit, the protons and 
antiprotons from the low $M_{\ppbar}$ region stop in the TOF counters
and, thus, have very characteristic BSC responses: protons do not
produce any matching signals in the BSC while secondary
particles from antiproton annihilation usually produce
large signals. This asymmetric behavior is quite distinct from that
for $K^+ K^-$, $\pipi$ or $\ee$ pairs,
where the positive and negative tracks produce similar, non-zero BSC
responses. 
The observed BSC energy distributions for the selected
$\jpsi\rt\gamma\ppbar$ events with $M_{\ppbar}\le 1.9$~GeV/$c^2$ 
closely match expectations for protons and antiprotons and show no
evidence for contamination from other particle species.

\begin{figure}[htb]
\centerline{\epsfysize 2.7 truein
\epsfbox{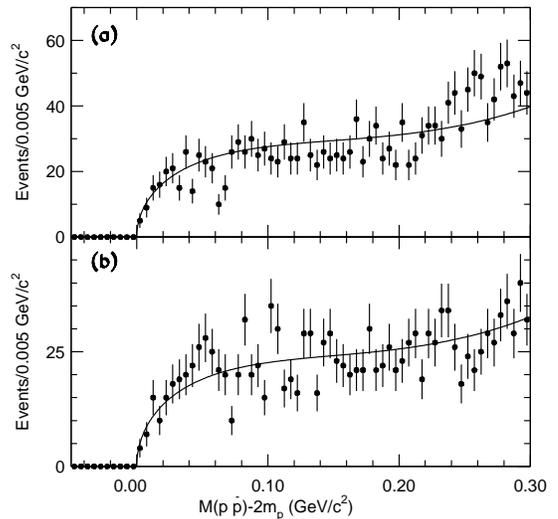}}
\caption{\label{fig:ppbar_mass}
The $M_{\ppbar}-2m_p$ distribution for {\bf (a)}
selected $\jpsi\rt\pi^0 \ppbar$ decays and
{\bf (b)} MC $\jpsi\rt\pi^0 \ppbar$ events that satisfy
the $\gamma\ppbar$ selection criteria. 
The smooth curves 
are the result of fits described in the text.}
\end{figure}

There is, however, a large background from $\jpsi\rt\pi^0\ppbar$ events
with an asymmetric $\pi^0\rt\gamma\gamma$ decay where
one of the photons has most of the $\pi^0$'s energy.  
This is studied using a sample of $\jpsi\rt\pi^0\ppbar$
decays reconstructed from the same data sample.  For
these, we select
events with oppositely charged tracks that are identified as protons
and with two or more photons, 
apply a four-constraint kinematic fit
to the hypothesis $\jpsi\rt\gamma\gamma\ppbar$, and
require $CL_{\gamma\gamma\ppbar}>0.005$. For
events with more than two $\gamma$'s, we select the $\gamma$ pair that
produces the best fit.  In the $M_{\gamma\gamma}$ distribution
of the selected events there is a
distinct $\pi^0$ signal; we require 
$\vert M_{\gamma\gamma}-M_{\pi^0}\vert < 0.03$~GeV/$c^2$ 
($\pm 2\sigma$).
The distribution of events {\it vs.} $M_{\ppbar} - 
2m_p$ near the $M_{\ppbar} = 2m_p$ threshold,
shown in Fig.~\ref{fig:ppbar_mass}(a), is reasonably
well described by a function of the form 
$f_{\rm bkg}(\delta) =N(\delta^{1/2}+ a_1 \delta^{3/2} + a_2
\delta^{5/2})$,
where $\delta \equiv M_{\ppbar} - 2m_p$ 
and the shape parameters
$a_1$ and $a_2$ are determined from a fit to simulated MC events
that were generated uniformly in phase space.   This is
shown in the figure as a smooth curve. There is no indication of a
narrow peak at low $\ppbar$ invariant masses. Monte Carlo simulations of
other $J/\psi$ decay processes with final-state $\ppbar$ pairs 
indicate that
backgrounds from processes other than $J/\psi\rt\pi^0
\ppbar$ are negligibly small.

The $M_{\ppbar}-2m_p$ distribution for the $\pi^0\ppbar$ phase-space MC
events that pass the $\gamma\ppbar$ selection
is shown in Fig.~\ref{fig:ppbar_mass}(b).
There is no  clustering at threshold;  the
smooth curve is the result of a fit to $f_{\rm bkg}(\delta)$ with 
the same shape parameter values.

In BESII, the detection efficiency for protons and antiprotons 
falls sharply for three-momenta below 0.4~GeV/$c$.
This produces a mass dependence in the experimental acceptance
near $M_{\ppbar}\simeq 2m_p$ for $\jpsi\rt\gamma\ppbar$ and 
$\pi^0\ppbar$.  For both processes, when $M_{\ppbar}$ is very near
$2m_p$, the $p$ and $\pbar$ both have three-momenta very near 0.5~GeV/$c$
and are well detected.  For increasing $\ppbar$ masses, more
asymmetric energy sharing is possible and the acceptance
decreases until $M_{\ppbar}\simeq 2.0$~GeV/$c^2$, where it is 
$\simeq 0.65$ of its value at $M_{\ppbar} = 2m_p.$  

\begin{figure}[htb]
\centerline{\epsfysize 2.9 truein
\epsfbox{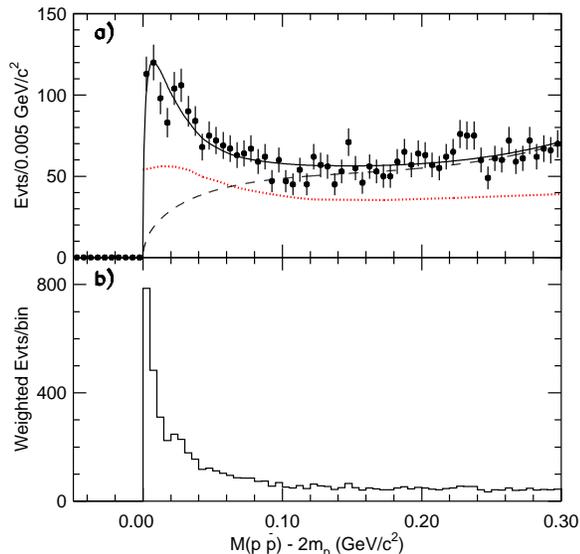}}
\caption{\label{fig:2pg_thresh_fit}
(a) The near threshold $M_{\ppbar}-2m_p$ distribution for
the $\gamma\ppbar$ event sample.  The dashed
curve is the background function described
in the text.  The dotted curve
indicates how the acceptance varies with $\ppbar$ invariant
mass; the dashed curve shows the fitted background function.
(b) The  $M_{\ppbar}-2m_p$ distribution with events weighted
by $q_0/q$.}
\end{figure}

Figure~\ref{fig:2pg_thresh_fit}(a) shows the threshold region
for the selected $\jpsi\rt\gamma\ppbar$ events.  
The dotted curve in the figure
indicates how the acceptance varies with invariant
mass.  The solid curve shows
the result of a fit using an 
acceptance-weighted $S$-wave Breit-Wigner (BW)
function~\cite{kcube}
to represent the low-mass enhancement plus
$f_{\rm bkg}(\delta)$ to represent the background.
The mass and width of the BW signal
function are allowed to vary and the shape parameters
of $f_{bkg}(\delta)$ are fixed at the values derived from the fit
to the $\pi^0\ppbar$ phase-space MC sample~\cite{resol}.
This fit yields $928\pm 57$ events
in the BW function with a peak mass of
$M=1859 ^{~+3}_{-10}$~MeV/$c^2$ and a full
width of $\Gamma = 0\pm 21$~MeV/$c^2$~\cite{fbkg}.  
Here the errors are statistical only:  those for the
event yield  and the width are derived from
the fit; the determination of the statistical
errors for the mass is discussed below.  The
fit confidence level is 46.2\% 
($\chi^2/d.o.f.= 56.3/56$).

Monte Carlo studies indicate that in the presence
of background, the determination of the peak mass
for a below-threshold resonance is more unreliable
the further the peak position is below threshold.
This produces an asymmetric distribution of mass input
values that can produce our measured result.  Moreover,
the rms spread of these values increases for lower
input masses, indicating that the statistical error
returned by our mass fit underestimates the negative
error.  Because of this, we quote statistical errors
for the mass that are derived from the rms spreads of
fit results for an ensemble of MC experiments with
different input mass values.

Further evidence that the peak mass is below the $2m_p$ 
threshold is provided in 
Fig.~\ref{fig:2pg_thresh_fit}(b), which shows the 
$M_{\ppbar}-2m_p$ 
distribution when the kinematic threshold behavior
is removed by weighting each event by $q_0/q$, 
where $q$ is the proton momentum in the $\ppbar$
restframe and $q_0$ is the value for $M_{\ppbar}=2$~GeV/$c^2$.
The sharp and monotonic increase at threshold
that is observed in this weighted histogram can only occur
for an $S$-wave BW function when the peak mass is below $2m_p$.

An $S$-wave $\ppbar$ system with even $C$-parity 
would correspond to a $0^{-+}$ pseudoscalar state.
We also tried to fit the signal with a $P$-wave BW function,
which would correspond to a $0^{++}$ 
($^3_0P$) scalar state that occurs in some models~\cite{Nambu,richard}. 
This fit yields a peak mass $M=1876.4 \pm 0.9$~MeV$c^2$, 
which is very nearly equal to $2m_p$, and a very narrow total
width: $\Gamma = 4.6 \pm 1.8$~MeV$c^2$ (statistical errors only).  
The fit quality, $\chi^2/d.o.f. = 59.0/56$, is worse 
than that for the $S$-wave BW but still acceptable.
A fit with a $D$-wave BW fails badly with
$\chi^2/d.o.f. = 1405/56$.

In addition we tried fits that use known particle resonances to 
represent the low-mass peak.  
There are two spin-zero resonances 
listed in the PDG tables in this mass region~\cite{PDG}: 
the $\eta(1760)$ with  $M_{\eta(1760)} = 1760 \pm 11$~MeV/$c^2$ and 
$\Gamma_{\eta(1760)} = 60 \pm 16$~MeV, and the $\pi(1800)$ with
$M_{\pi(1800)} = 1801 \pm 13$~MeV/$c^2$ and 
$\Gamma_{\pi(1800)} = 210 \pm
15$~MeV. A fit with $f_{\rm bkg}$ 
and an acceptance-weighted $S$-wave BW function with mass
and width fixed at the PDG values for the $\eta(1760)$ produces
$\chi^2/d.o.f. = 323.4/58$.  
A fit using a BW with the $\pi(1800)$ parameters
is worse.

\begin{figure}[htb]
\centerline{\epsfysize 1.5 truein
\epsfbox{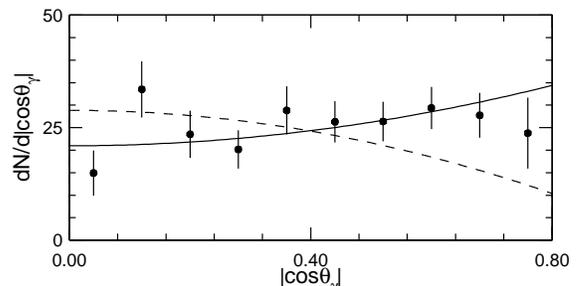}}
\caption{\label{fig:2pg_cosg_corr}
The background-subtracted, acceptance-corrected
$|\cos\theta_{\gamma} |$ distribution for
$\jpsi\rt\gamma\ppbar$-enriched events with $M_{\ppbar}\le
1.9$~GeV/$c^2$.  The solid curve is a fit to
a $1+\cos^2\theta_{\gamma}$ shape
for the region $|\cos\theta_{\gamma}|\le 0.8$;
the dashed curve is the result of a 
fit to $\sin^2\theta_{\gamma}$.}
\end{figure}

For both the scalar or pseudoscalar 
case, the  polar angle of the photon,
$\theta_{\gamma}$, would be distributed
according to $1 + \cos^2\theta_{\gamma}$.
Figure~\ref{fig:2pg_cosg_corr} shows the
background-subtracted,
acceptance-corrected $|\cos\theta_{\gamma}|$ 
distribution for events with
$M_{\ppbar}\le 1.9$~GeV and $|\cos\theta_{\gamma}|\le 0.8$.  
Here we have subtracted the $\vert \cos\theta_{\pi^0}\vert $
distribution from the $\pi^0\ppbar$ data sample, normalized to
the area of $f_{\rm bkg}(\delta)$ for $M_{\ppbar}<1.9$~GeV/$c^2$ to
account for background. The
solid curve shows the result of a fit for $1 + \cos^2\theta_{\gamma}$ to
the $\vert\cos\theta_{\gamma}\vert<0.8$ region;
the dashed line shows the result of a 
similar fit to $\sin^2\theta_{\gamma}$.  
Although the data are not precise enough to establish a 
$1 + \cos^2\theta_{\gamma}$ behavior, the
distribution is consistent
with expectations for a radiative transition
to a pseudoscalar or scalar meson~\cite{angle}.

We evaluate systematic errors on the mass and width
from changes observed in the fitted values for
fits with different bin sizes, with background shape 
parameters left as free parameters, different shapes
for the acceptance variation, and different resolutions.  
The ensemble Monte Carlo studies mentioned
above indicate that in the presence of
background, the determination of the parameters
of a sub-threshold BW resonance can be biased.  We include the
range of differences between input and output values seen
in the MC study in the systematic errors.

For the mass, we determine a systematic error 
of $^{~+5}_{-25}$~MeV$c^2$.
For the total width, we determine a 90\% confidence
level (CL) upper limit
of $\Gamma < 30$~MeV/$c^2$, where the limit includes
the systematic error.

Using a Monte-Carlo determined acceptance of $23\%$, we determine
a product of branching fractions ${\cal B}(\jpsi\rt\gamma X(1859))
{\cal B}(X(1859)\rt\ppbar) = (7.0 \pm 0.4 {\rm (stat)} 
^{+1.9}_{-0.8}{\rm (syst)})\times
10^{-5}$, where the systematic error includes uncertainties in the
acceptance (10\%), the total number of $\jpsi$ decays in the 
data sample (5\%), and the effects of changing the various inputs to the 
fit ($^{+24\%}_{~-2\%}$).

In summary, we observe a strong, near-threshold enhancement in the
$\ppbar$ invariant mass distribution in the radiative decay
process $\jpsi\rt\gamma\ppbar$. No similar structure is seen 
in $\jpsi\rt \pi^0\ppbar$ decays.
The structure has properties consistent with either a $J^{PC} = 0^{-+}$
or $0^{++}$ quantum number assignment and cannot be attributed
to the effects of any known meson resonance.  If interpreted as a 
single $0^{-+}$ resonance,
its peak mass is below the $M_{\ppbar}=2m_p$ threshold at
$1859 ^{~+3}_{-10} {\rm (stat)} ^{~+5}_{-25} {\rm (syst)}$ MeV/$c^2$ and
its width is $\Gamma < 30 $~MeV/$c^2$ at the 90\% CL.

We thank the staffs of BEPC and the computing center
for their strong efforts.
This work is supported in part by the National Natural Science Foundation
of China under contracts Nos. 19991480, 10225524, 10225525 and the 
Chinese 
Academy of Sciences under contract No. KJ 95T-03, the 100 Talents Program 
of CAS under Contract Nos. U-24, U-25, and the Knowledge Innovation 
Project of CAS under Contract Nos. U-602, U-34 (IHEP); by the National 
Natural 
Science 
Foundation of China under Contract No.10175060(USTC); and
by the Department of Energy under Contract Nos.
DE-FG03-93ER40788 (Colorado State University),
DE-AC03-76SF00515 (SLAC), DE-FG03-94ER40833 (U Hawaii), 
DE-FG03-95ER40925 (UT Dallas).

%\end{thebibliography}

\begin{references}
\bibitem{kcube} For the BW, we use the form:
\[
BW(M) \propto \frac{q^{(2\ell+1)} k^3}{(M^2-M_0^2)^2 + M_0^2\Gamma^2},
\]
where $\Gamma$ is a constant (determined from the fit), 
$q$ is the proton momentum in the $\ppbar$ restframe,
$\ell$ is the $\ppbar$ orbital angular momentum, and $k$ is the 
photon momentum.
\bibitem{resol} 
The $\ppbar$ mass
resolution 
varies from $\sigma\simeq 1.2$~MeV/$c^2$
at $M_{\ppbar}\simeq 2m_p$,
to $\sim 3$~MeV/$c^2$ at higher
masses.  Convolving the fitting function with
a Gaussian with a width in this range
has no significant effect on the results.
\bibitem{fbkg} 
The background level in the fit is about twice the
level of $\jpsi\rt\pi^0\ppbar$ feedthrough predicted by
the MC.  This indicates that there is some contribution 
from nonresonant three-body $\jpsi\rt\gamma\ppbar$ decays.
\bibitem{angle} 
The $1+\cos^2\theta_{\gamma}$ fit has $\chi^2/d.o.f. = 3.7/7$; the
$\sin^2\theta_{\gamma}$ fit has $\chi^2/d.o.f. = 10.3/7$.
\end{references}
\end{document}